\documentclass[12pt]{article}
\DeclareUnicodeCharacter{2212}{\ensuremath{-}}
\usepackage[utf8]{inputenc}
\usepackage{amsmath}
\usepackage{amsfonts}
\usepackage{amssymb}
\usepackage{graphicx}
\usepackage{geometry}
\usepackage{hyperref}
\usepackage{multirow}
\usepackage[english]{babel}
\usepackage[bottom]{footmisc}
\usepackage{graphicx,algorithmic}
\usepackage[linesnumbered,ruled,vlined]{algorithm2e}
\SetCommentSty{mycommfont}

\SetKwInput{KwInput}{Input}                
\SetKwInput{KwOutput}{Output}              
\title{Unraveling implicit human behavioral effects on dynamic characteristics of Covid-19 daily infection rates in Taiwan}
\author{T-L, Chen\footnotemark[1], P-T, E. Chou\footnotemark[2], M-Y, Chen\footnotemark[2] and H. Fushing\footnotemark[3] \footnotemark[4]}
\date{\today}
\begin{document}
\maketitle
\footnotetext[1]{Institute of Statistical Science, Academia Sinica, Taipei, Taiwan}
\footnotetext[2]{Department of Statistics, National Chengchi University, Taiwan}
\footnotetext[3]{Department of Statistics, University of California at Davis, CA, 95616.}
\footnotetext[4]{Correspondence: Department of Statistics, University of California at Davis, E-mail: fhsieh@ucdavis.edu}

\begin{abstract}
We study Covid-19 spreading dynamics underlying 84 curves of daily Covid-19 infection rates pertaining to 84 districts belonging to the largest seven cities in Taiwan during her pristine surge period. Our computational developments begin with selecting and extracting 18 features from each smoothed district-specific curve. This step of computing effort allows unstructured data to be converted into structured data, with which we then demonstrate asymmetric growth and decline dynamics among all involved curves. Specifically, based on Theoretical Information measurements of conditional entropy and mutual information, we compute major factors of order-1 and order-2 that reveal significant effects on affecting the curves' peak value and curvature at peak, which are two essential features characterizing all the curves. Further, we investigate and demonstrate major factors determining the geographic and social-economic induced behavioral effects by encoding each of these 84 districts with two binary characteristics: North-vs-South and Unban-vs-suburban. Furthermore, based on this data-driven knowledge on the district scale, we go on to study fine-scale behavioral effects on infectious disease spreading through similarity among 96 age-group-specific curves of daily infection rate within 12 urban districts of Taipei and12 suburban districts of New Taipei City, which counts for almost one-quarter of the island nation's total population. We conclude that human living, traveling, and working behaviors do implicitly affect the spreading dynamics of Covid-19 across Taiwan profoundly.
\end{abstract}

\section{Introduction}
As a relatively new public health crisis threatening the whole world, the emergence and spread of the coronavirus disease 2019 (COVID-19) pandemic, which cause the novel severe acute respiratory syndrome coronavirus 2 (SARS-CoV-2), has caused many countries worldwide to suffer from sudden and overwhelming burdens in hospitalizations for pneumonia with multiorgan disease and many millions of death \cite{wiersinga,singhal}. For the past three years, this pandemic has also caused uncountable economic losses to almost all countries as well as psychological and financial losses to billions of individuals \cite{cheng,elgin}. However, the spreading dynamics of this infectious disease beyond settings of a private home, a crowdy restaurant, or a concert hall, such as districts in a city, is still, by and large, missing in Covid-19 literature. This particular scale seems to have some degrees of homogeneity from geographic and social-economic perspectives compared with a whole city or country scale. While several works have studied this scale of spread dynamics, they all focused on evaluating model-based effects of emergency containment measures, and non-pharmaceutical interventions \cite{gatto,schuppert}, instead of its intrinsic nature. The lacking of the data-driven intrinsic nature of spreading dynamics indeed signals the knowledge gap in understanding linking the infectious disease's spreading dynamics to human geographic and social-economic dynamics and vice versa.

To fill this knowledge gap in Covid-19 literature, instead of discussing medical, economic, psychological, and financial aspects of the Covid-19 crisis, this paper studies its spreading dynamics within the island nation Taiwan. One unique characteristic role Taiwan plays in the pandemic is that it can almost be taken as a close domain for this infectious disease. Hence, its spreading dynamics are worth looking into.

To unravel the spreading dynamics of Covid-19 in Taiwan, we begin with a unique story that has been taking place in this island nation. Our motivation for this story rests purely on highlighting the importance of uncharacterized and unobserved human behaviors that could have implicitly driven this pandemic crisis across the entire geographic and social constituents. From the beginning of the Covid-19 pandemic crisis at the end of the year 2019 to March 2022, the entire Taiwanese population of 23 million has been subject to very strict regulating measures on social gatherings and a very thorough contact-tracing protocol on movements of infected individuals. Under strict measures and thorough contact tracing, the numbers of total infection cases and related death have been kept amazingly low. Consequently, almost all daily activities go on as usual for most of 30 months period. During this period, except for no outgoing international travel and incoming international visitors, it is fair to say that residents in Taiwan tend to think as if the Covid-19 pandemic crisis has been going on only in the outside world.

Within this more than a two-year period, many insurance companies in Taiwan have come to speculate on the potential market for personal Covid-19 insurance policies. Many insurance products were pushed out, but the sales were not so great because no great incentives existed. Not until April 21, 2022, the Central Epidemic Command Center (CECC), which is responsible for all regulating policies regarding this pandemic, switched its principle from ``completely blocking'' to ``steadily co-existing''. Then CECC started to modify regulating measures toward the relaxing direction, including quarantine requirements, and to give up the thorough contact-tracing protocol completely because it had broken down when facing several hundreds of daily infected cases. CECC's policy changes also reflected the fact that high percentages of the population, except for children under 12 years old, have been vaccinated with at least two doses. To this date, the number of insurance policy purchases is about 1.91 million, which is more than 8$\%$ of the entire population.

\begin{figure}
 \centering
   \includegraphics[width=6in]{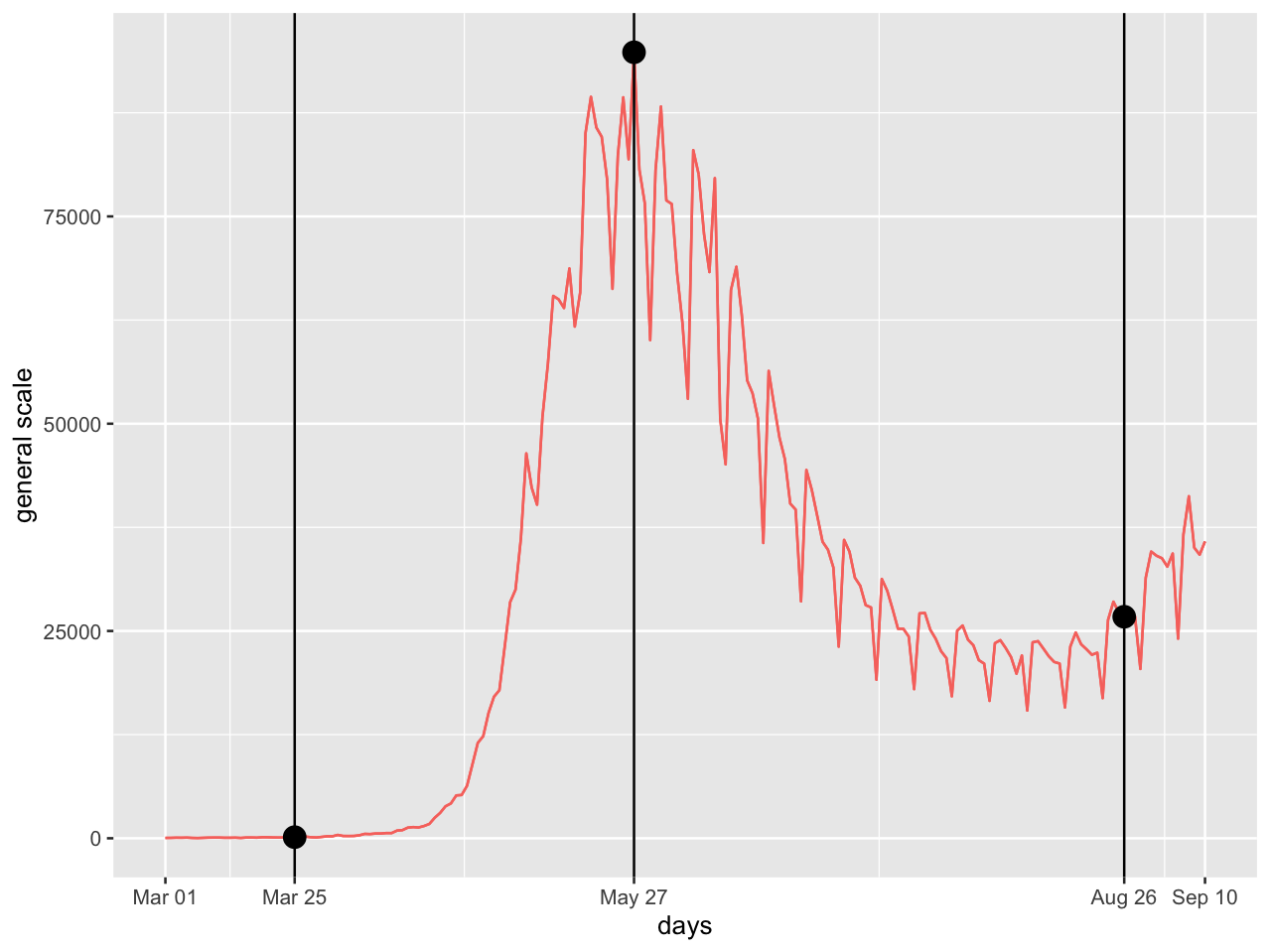}
 \caption{Taiwan's curve of daily infection rate from March to September 2022.}
 \label{Taiwancovid}
 \end{figure}

As shown in Fig.~\ref{Taiwancovid}, the curve of the daily infection rate in Taiwan surged right up after April 21, 2022. People then realized the number of cases of Covid-19 infection was expected to go up quickly, and the spreading of infectious disease would be in a whole scale fashion across Taiwan. People then rushed to buy Covid-19 insurance policies. At the same time, this rush was hugely aided by a government policy change on allowing claims on quarantine-at-home as hospitalization due to Covid-19 infection. Even though this policy change also led five insurance companies to stop selling all their Covid-19 insurance products on April 26, 2022, this rush of buying Covid-19 insurance products from the rest of the 15 insurance companies continued to grow sharply.

\begin{figure}
 \centering
   \includegraphics[width=6in]{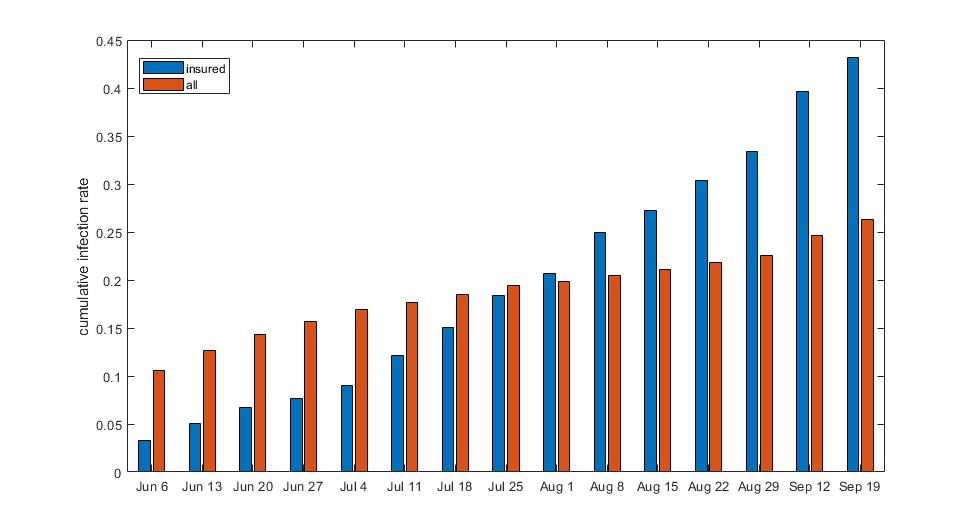}
 \caption{The comparison of the cumulative infection rates between the insured group and the entire population in Taiwan.}
 \label{Insurcovid}
 \end{figure}

From Fig.~\ref{Insurcovid}, we can see the weekly comparison of two cumulative infection rates: the insured group and the general public. We use the odds ratio to reveal a glimpse of the infection difference between these two groups. By August 31, there were more than 4.714 million policies sold. Among them, there are about 1.573 million claims have been made. To make a comparison with the general public, we give a contingency table given as follows:
\begin{table}[h!]
\centering
\begin{tabular}{lll}\hline
Insurance status/Covid-19 status & 0& 1\\ \hline
1&3.142 &1.573\\
0&14.741&3.735\\\hline
\end{tabular}%
\caption{Contingency table of insurance status-vs-Covid-19 status. Cell counts in millions.}
\label{Insurance}
\end{table}
The odds of being infected in the insured group is $0.5008=\frac{1574}{3142}$, while the odds of being infected in the non-insured group is $0.2534=\frac{3735}{14741}$. So the odds ratio of being infected by Covid-19 in the insured group relative to the non-insured group is 1.9765. In fact, the odds ratio on September 19 is calculated as high as 2.7000. That is, people who bought Covid-19 insurance are more than twice as likely to get infected than those who did not buy Covid-19 insurance. The ending of this story is that these Taiwanese insurance companies collectively lost about 3 billion USD in September. An even more prominent figure is expected in October 2022.

This simple story is to reiterate the well-known fact that Covid-19 infection dynamics within a population are fundamentally related to human behaviors. However, this fact is sharply confronted with another fact: No human behavioral data of the entire population could be publicly available. Since the entire population is susceptible to this infectious disease, Collecting behavioral data from 23 million is not yet possible at this stage. Therefore, we need to extract behavior-related information from geographic locations and social-economic statuses.

 Since cities in northern Taiwan are more densely populated than cities in southern Taiwan, traveling from the North-to-South is indeed taken as a way of lessening the likelihood of Covid-19 infection. By encoding the cities of the North along Taiwan's high-speed railway against those of the South, we can implicitly bring out potential behavioral effects induced by geographic differences.

Taipei is a city that houses a majority of government branches and large business cooperations. Its real estate price is in general higher than any other city because of better educational and medical systems among many other aspects. That is, Taipei has better social-economic status than all other cities surrounding it. By encoding Taipei as urban and its surrounding city as suburban, we can implicitly bring out potential behavioral effects induced by social-economic differences. Therefore, we indeed do not need to collect individuals' behavioral data in order to evaluate potential behavioral effects on Covid-19 spreading dynamics. This is our approach to achieving one of the two major goals of this paper.

The other major goal of this paper is to propose and develop a simple computational framework for exploring the spreading dynamics of infectious diseases. This goal is motivated to fill the gap of lacking analytical methodologies for studying population-wise spreading dynamics of infectious diseases based on curves of daily infection rates coming out of many administrative districts. The unstructured curve data type surely poses a challenge in data analysis.
All district-specific curves of daily infection rates used in this paper are downloaded from the official website of the Central Epidemic Command Center (CECC) of Taiwan.

The study period of this research begins on March 25 and ends on August 19, 2022. The ending date is at the tough of the overall curve of daily infection rate before the arrival of another surge of Omicron BA.5 variant, as seen in Fig~\ref{Taiwancovid}. It is noted that, within Fig~\ref{Taiwancovid}, the date of August 26 is marked just because data of August 26 is involved in the moving average at August 19.

Within this study period, we aren't concerned about potential reinfection issues. Since, before the arrival of Omicron, it is estimated that only one in a hundred could be reinfected within 6 months \cite{medic}. And we only focus on daily Covid-19 infection rates without going into any medical aspects of this disease, such as Long-COVID among many others \cite{taquet}.

Both major goals of this study are to help general as well as scientists understand what are key factors driving the fundamental characteristics of dynamics underlying the curve of daily infection rate in a data-driven fashion. We are particularly interested in patterns related to growth to peak-value and curvature-at-peak to decline (something not clear). Since pertaining to a city, a district in a city, or even an age group within a district of a city, each curve of daily infection rate is of the unstructured data type or so-called functional data. It is theoretically infinite-dimensional. In this study, we extract 18 measurable features from each curve. These 18 features selected and extracted in this paper are designed to coherently characterize the growth and decline patterns embedded within each curve. That is, such a curve is then represented by an 18-dim vector of measurements of these 18 features. This vector data form is of the so-called structured data type.

We then apply the major factor selection protocol to respectively compute two collections of major factors that can collectively explain individual dynamics of response features: peak-value and curvature-at-peak, by having significantly high associations. Here, a major factor is a feature set that can explain a large proportion of uncertainty of the response variable. This data-driven computational protocol has been developed in a series of works \cite{CCF22a,CCF22b,FCC23a,FCC23b} by employing Theoretical Information Measurements, such as conditional entropy and mutual information \cite{cover}. Such data-driven patterns related to the spreading dynamics of infectious diseases are relatively new in the literature.

In this study, Taipei is taken as an urban collective of 12 districts, while New-Taipei-city is taken as a suburban collective of 12 selected districts. These two cities accommodate more than 6 million people which counts for at least one-quarter of Taiwan's total population. We also include 5 more cities: two are encoded as suburban because of being directly connected to New Taipei City and the rest 3 are located in the south of New Taipei City. Within each of these three southern cities, districts are classified as urban against suburban according to their  registration statuses before merging. Further, we investigate fine-scale of the dynamics of Covid-19 spreading through similarity and dissimilarity across 96 different age groups belonging to 12 districts of Taipei and New Taipei City.

By focusing on Covid-19's spreading dynamics, we hope to provide a different analytic toolset for researchers in the field of epidemiology. In particular, we like to bring out the intrinsic nature hidden in this infectious disease. By evaluating behavioral effects on spread dynamics, we hope to provide an extra data analysis toolset to researchers in social sciences and economic policy to take human behaviors into consideration as well as to motivate them to use data-driven approaches beyond standard statistics packages \cite{cheng,elgin}.

\section{Feature extraction from curves of daily infection rate.}
Given that a curve of daily infection rate pertaining to an administrative district is unstructured data, like a continuous curve or an image, its relevant pieces of information regarding to growth and decline would remain hidden before being successfully extracted. In this section, we discuss how to extract such relevant pieces of information from each curve of daily infection rate. This information extraction procedure is rather intuitive as given below. The procedure begins by identifying measurable characteristics of an unstructured curve. Each measurable characteristic is computed and then termed as a measurement of a characteristic-specific feature. That is, one piece of the curve's information is encoded in a measurement of a feature. The resultant set of selected features amounts to collectively capture the curve's intrinsic information.

\begin{figure}
 \centering
   \includegraphics[width=6in]{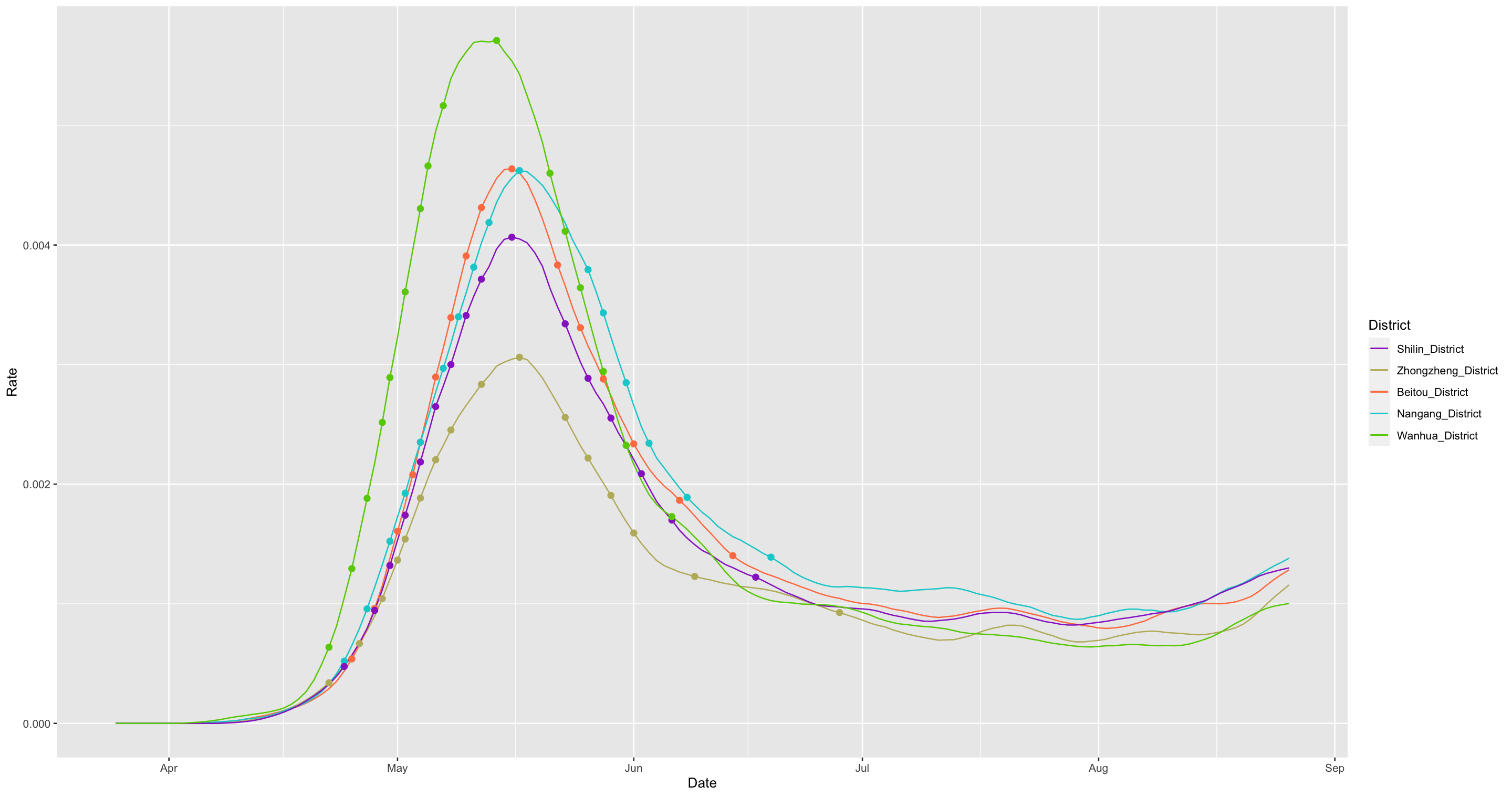}
 \caption{Five smoothed curves of Taipei's 12 districts}
 \label{5fromtaipei}
 \end{figure}

At first, we generically represent a curve of daily infection rate in the form of time series as $X_t$ along the day-axis of $t$. Our pre-processing step begins with smoothing the curve by taking the 7-day moving average twice or equivalently by performing the moving 13-day weighted average with respect to a specified weighting scheme: denoting the smoothed curve as $\tilde{X}_t$,
\[
\tilde{X}_t=\sum^6_{d=-6} (\frac{7+d}{49})\times X_{t+d}.
\]
The reason behind this specific weighting scheme is to fully accommodate the somehow persistent weekly pattern of daily reported infected cases as seen in Fig.~\ref{Taiwancovid}. That is, the needed smoothness can only be achieved by overcoming the weekly pattern. Our empirical observations indicate that the trajectory of $\tilde{X}_t$, in general, would first strictly grow until the unique peak and then steadily decline until the end of the study period on August 19. An illustrative figure Fig~\ref{5fromtaipei} demonstrates 5 smoothed curves pertaining to 5 districts of Taipei. Each of these five smoothed curves of daily infection rate has one unique mode. On the left of the peak, the growth is sharp, while, on the right of the peak, the decline is more gentle. We take such uni-mode, sharp growth, and gentle decline patterns as a curve's defining characteristics.  In this section, our features are to be selected and extract to collectively embrace such characteristic patterns.

Based on the smoothed curve $\tilde{X}_t$, we first extract two peak-related features: 1) the date of ``peak'' $t_{max}$ as the day-$t$ when $\tilde{X}_t$ achieves its maximum value; 2) and its ``peakvalue'' $\tilde{X}_{t_{max}}$ . Then, we extract the first time when $\tilde{X}_t$ grows and passes the $90\%$ of $\tilde{X}_{t_{max}}$ before reaching $t_{max}$. Denote this feature as $t_{-0.1}$ with a specification given below as:
\[
t_{-0.1}= \inf \{t | \tilde{X}_t \geq 0.9 \times \tilde{X}_{t_{max}} \}.
\]
The negative sign in the subscript is to indicate being on the left of $t_{max}$. In contrast, we define $t_{0.1}$ on the right of $t_{max}$ as the last time after $t_{max}$ when $\tilde{X}_t$ remains to stay above the $90\%$ of $\tilde{X}_{t_0}$:
\[
t_{0.1}= \inf \{t > t_{max}| \sup_{t_{max} < s \leq t} \tilde{X}_s \geq 0.9 \times \tilde{X}_{t_{max}} \}.
\]

Likewise, we extract 7 features $\{t_{-0.8}, t_{-0.7},.., t_{-0.2} \}$ on the left of $t_{-0.1}$ and another 7 features $\{t_{0.2}, t_{0.3},.., t_{0.8} \}$ on the right of $t_{max}$ as well. Their precise definitions as given as follows: for $0.2\leq \alpha \leq 0.8$,
\begin{eqnarray*}
t_{-\alpha}&=& \inf \{t < t_{max} | \tilde{X}_t \geq (1-\alpha) \times \tilde{X}_{t_{max}} \},\\
t_{\alpha}&=& \inf \{t >t_{max} | \sup_{t_{max} < s \leq t} \tilde{X}_s \geq (1-\alpha) \times \tilde{X}_{t_{max}} \}.
\end{eqnarray*}

With the information of these 18 features derived from $\tilde{X}_t$, we can more or less reconstruct the $\tilde{X}_t$ with relatively high precision. For instance, we can evaluate the positive slope of $\tilde{X}_t$ on interval $(t_{-0.20}, t_{-0.30})$ and negative slope of $\tilde{X}_t$ on interval $(t_{0.30}, t_{0.20})$, etc. However, we do not extract $t_{-0.10}$ due to the stability issue, while $t_{0.1}$ has not yet been reached by any districts.

\begin{figure}
 \centering
   \includegraphics[width=6in]{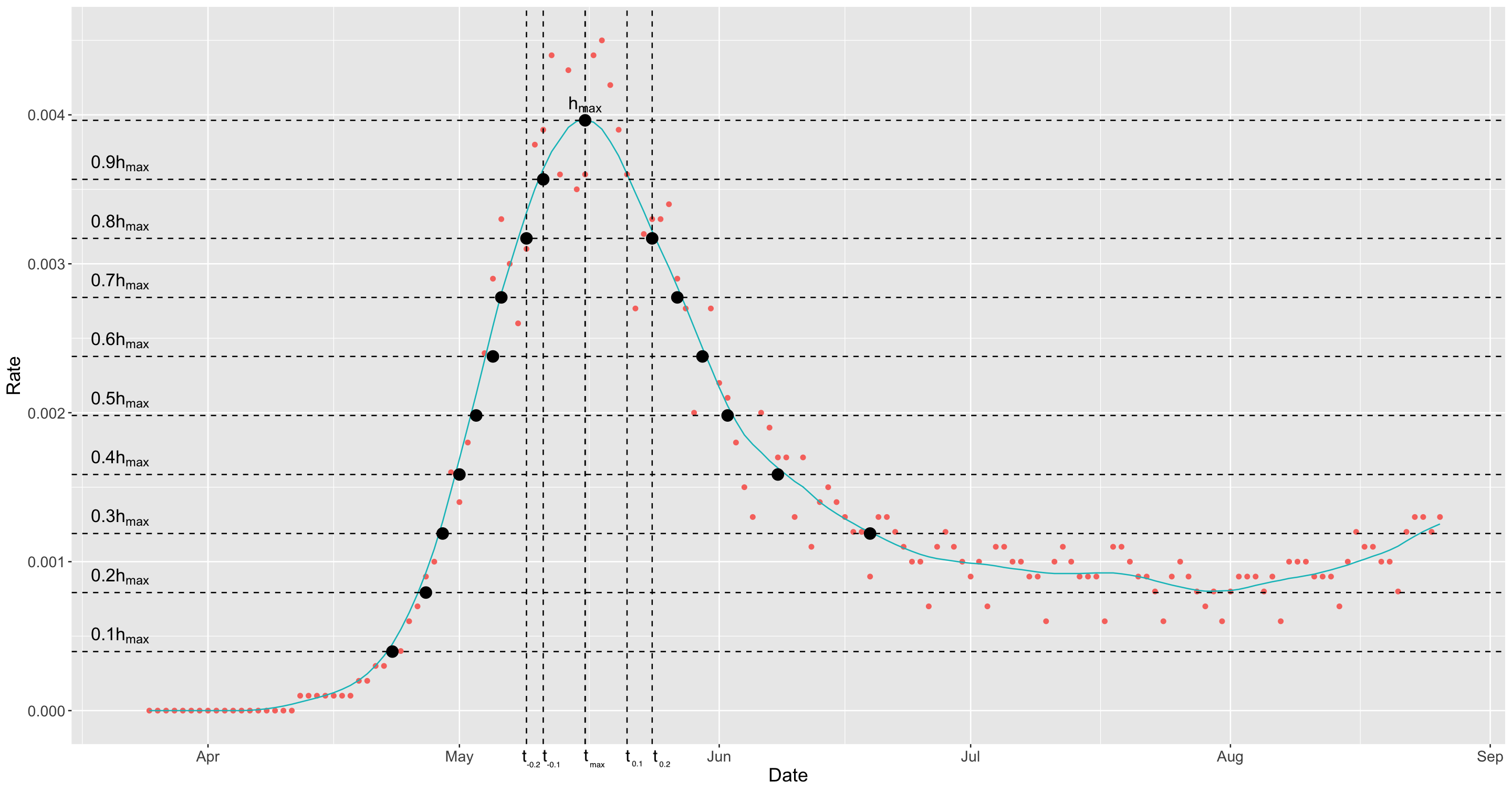}
 \caption{Illustrative definitions and extractions of all features}
 \label{allfeature}
 \end{figure}

On top of the above 18 features, we also derive two calculated features as follows. The first calculated feature is defined as the so-called robust-peak $t_0$ as the middle point, rounded down to an integer, of the interval $[t_{-0.1}, t_{0.1}]$. In most districts, $t_0$ and $t_{max}$ are almost equal. It is evident that $t_0 > t_{max}$ due to asymmetry of growth and decline generally observed across all curves of $\tilde{X}_t$. The second calculated feature is the so-called ``curvature-at-peak'' defined as the length of this interval $[t_{-0.1}, t_{0.1}]$, which to a great extent indicates the curvature of $\tilde{X}_t$ at $t_0$. The two calculated features are to be used to define response variables within investigations for characterizing dynamics underlying the formation of curves of daily infection rates. See figurative illustrations of all these defined quantities in Fig~\ref{allfeature}.

To enable us to compare the shapes of all curves of daily infection rates and retain their semantic clarity, we define the following variables based on the above-defined 18 features. The feature-variable of ``peak'' is extracted as $t_{max}-t_{0}$,  ``peakvalue '' as $\tilde{X}_{t_{max}}$, ``left90'' as $t_{0}-t_{-0.90}$, ``right90'' as $t_{0.90}-t_{0}$, ``left80'' as $t_{0}-t_{-0.80}$, ``right80'' as $t_{0.80}-t_{0}$, and so on. That is, by doing so, we remove the calendar coordinate information out of all features $t_{-\alpha}$ and $t_{\alpha}$ for all $0.2\leq \alpha \leq 0.8$ and only retain their shape information. Henceforth, we can compare shapes of two curves of daily infection rate via their values of 18 feature-variables: \{peak, peakvalue, left90, left80,...,left20, right90, right80,..., right20.\} as if they are aligned with respect to one common $t_{0}$. The only feature variable that retains the calendar coordinate is the feature $t_{max}$ for ``peakdate''.

Measurements of these 18 feature variables across all potential district-specific or age-group-specific curves are collected from the largest 7 cities under study in Taiwan. Such measurements are of discrete data type. And all 18 dim vectors are stored in a structured data matrix format. These feature variables certainly retain their mutual associations of various degrees. In fact, the variations among degrees of the association are expected to be significantly heterogeneous. For instance, those feature variables on the Left depicting growth patterns are expected to be highly associative, while those feature variables on the Right depicting decline patterns are less so. Such associative asymmetry indeed reveals their essential roles among many other distinct roles in characterizing dynamics underlying curves of daily infection rates in the next section.

\section{Methods}
In this section, based on the structured data matrix constituted by the measurements of the 18 selected features, we will briefly introduce computational concepts and methodologies based on Theoretical Information Measurement (\cite{cover}) needed for the data analysis reported in this paper. In the first subsection, we review the concepts of conditional entropy and mutual information. The conditional entropy concept would be used to construct directed and undirected association measurements between two features as well as two feature sets of different sizes. In the second subsection, we briefly review the major factor selection protocol used to investigate the dynamics of a chosen response variable against a set of covariate features. This protocol has been developed in \cite{CCF22a,CCF22b,FCC23a,FCC23b} based on the concepts of conditional entropy and mutual information.
\subsection{Association based on Theoretic Information}
Many different statistics can measure the association between two variables. In this paper, we will analyze it based on Theoretic Infomation. The entropy of a variable $X$, defined as $H(X)={\mathbb E}_p(-\log(p(X)))$, represents the information or uncertainty of $X$. To investigate the association between two variables $X$ and $Y$, we can look into $X|Y$, $Y|X$, or $(X, Y)$. It can show that
\begin{eqnarray*}
    H(X,Y)&=& H(X) + H(Y|X) \\
    &=& H(X|Y) + H(Y)\\
    &=& H(X|Y) + I(X,Y) + H(Y|X),
\end{eqnarray*}
where
\[I(X,Y)=H(X)-H(X|Y)=H(Y)-H(Y|X)\]
is the mutual information of $X$ and $Y$. $H(X|Y)$ represents the uncertainty of $X$ after knowing $Y$ or the information of $X$ not explained by $Y$. We define the re-scaled conditional entropy
\[
{\cal E}(X|Y) = \frac {H(X|Y)} {H(X)}
\]
to measure the ratio of the information of $X$ not explained by $Y$. Note that $0 \leq {\cal E}(X|Y) \leq 1$.  When ${\cal E}(X|Y) =0$, $Y$ can perfectly explain $X$. On the contrary, ${\cal E}(X|Y) =1$ indicates that knowing $Y$ does not help understand $X$.

Based on this statistic, we build a network graph of all variables. Each variable is a vertex, and an edge connects two variables if the re-scaled conditional entropy is smaller than a fixed threshold. Furthermore, we use the edge width to indicate the association between the two variables.

In addition to the network presentation, we also illustrate the association of variables by a heatmap. The heatmap shows the re-scaled conditional entropies between variables. This heatmap's rows and columns are rearranged based on the corresponding Hierarchical Clustering Tree. In this way, the structures of the variables can be clearly revealed.

\subsection{Major factor selection}
As discussed in the previous subsection, $H(Y|X)$ indicates the information of $Y$ not yet explained by $X$. When $Y$ is a response variable, the lower $H(Y|X)$ is, the more likely $X$ is an essential factor to explain $Y$. For a given response variable $Y$, our goal is to find a subset ${\cal F} \subset \{X_1, X_2, \ldots, X_p\}$ such that $H(Y|{\cal F})$ is minimized. ${\cal F}$ is the set of feature variables to explain $Y$, which we name major factors of the response-versus-covariate (Re-Co) dynamics.

In \cite{CCF22a}, we proposed a protocol for major factor selection. The algorithm starts with investigating the 1-feature effect by calculating the entropies conditional on every single covariate. The covariates with significantly lower conditional entropy are potential candidates for the major factors of this Re-Co dynamics.

Next, we compute the conditional entropies on every pair of covariates.
In theory, $H(Y|X_i) \geq H(Y|X_i,X_j)$, and the inequality is usually strict due to the finite sample phenomenon. Therefore, it seems that it is usually better to add another covariate into a feature set. However, it is reasonable to ask for the covariate newly added to be significantly better than pure independent noise. We provide a fast algorithm to estimate the entropy reduced by noise, which is a standard to test whether the entropy dropped is significant.

In addition to checking whether there is a redundant covariate in a feature set, we also investigate the 2-order effect, which only exists if both covariates are presented. That is the case when the 2-feature effect is much larger than the sum of both 1-feature effects. The 2-feature effect is $H(Y)-H(Y|X_i,X_j)$, and the 1-features effects are $H(Y)-H(Y|X_i)$ and $H(Y)-H(Y|X_j)$. To check $$H(Y)-H(Y|X_i,X_j) \gg H(Y)-H(Y|X_i) +H(Y)-H(Y|X_j), $$
it suffices to check $$H(Y | X_i)-H(Y|X_i,X_j) \gg  H(Y)-H(Y|X_j). $$
Successive conditional entropy (SCE) drop is defined as
\[
\mbox{SCE-drop}({\cal F}) = \min_{X_i \in {\cal F}} H(Y| {\cal F} \setminus X_i) - H(Y|{\cal F}).
\]
Therefore, we check whether the 2-order effect exists from
\[
\mbox{SCE-drop}(X_i,X_j) \gg \min\{\mbox{SCE-drop}(X_i),\mbox{SCE-drop}(X_j) \}.
\]
Higher-order effects are checked by similar formulas.

\section{Asymmetric associative relationships among all feature-variables}
As demonstrated in Fig~\ref{allfeature}, these 18 feature variables are designed to capture the growth and decline patterns pertaining to curves of the daily infection rate over the fixed study period.  We explicitly evaluate and display the associative relationships among these 18 feature variables. Intuitively, such associations are revealed through their common shape as shown in Fig~\ref{5fromtaipei}. In this section, we employ conditional entropy and mutual information measurements to explicitly assess their degrees of association.

We first categorize each of the 18 feature variables into categorical ones with 4 categories. Such a categorization not only achieves noise reduction but also facilitates discoveries of potentially nonlinear associative patterns via the platform of the contingency table. That is, we can build a $4\times 4$ contingency table for any pair of these 18 features. Upon such a contingency platform, as discussed in the Method section, conditional entropy is employed respectively to evaluate row-variable-to-column-variable and column-variable-to-row-variable directional associations. By averaging these two directional associations, a mutual conditional entropy is obtained as the pair's nondirectional association.

In the computations of row-variable-to-column-variable association, each row of such a $4\times 4$ contingency table is used to identify a multinomial random variable specified by the row-sum and the vector of proportions, which is a vector of probability that gives rise to a row-wise conditional (Shannon) entropy. Usually, such a row-wise conditional entropy is re-scaled with respect to the entropy of the vector of column-sum proportions. This re-scaled conditional entropy reveals the proportion of uncertainty inside the column feature variable being reduced. This reduction is attributed to the information on the row feature variable's category.

\begin{figure}
 \centering
   \includegraphics[width=6in]{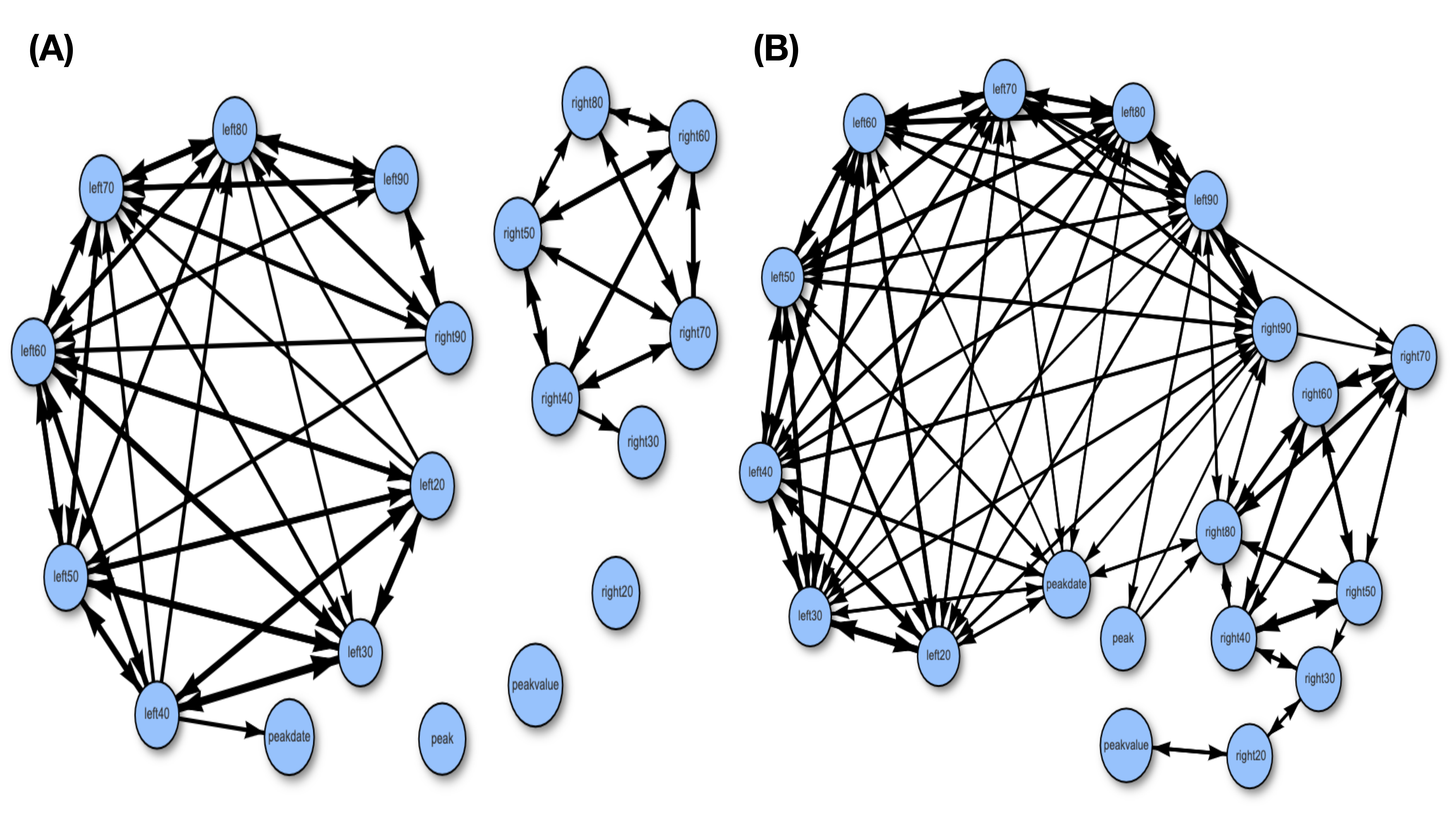}
 \caption{Directed associative network of all features with thresholding at: (A) 0.6; (B) 0.7}
 \label{dirnetwork}
 \end{figure}

The row-to-column directional association is the weighted mean of the four re-scaled conditional entropies. Similarly, we calculate the column-to-row directional association. Two versions of networks of directed associations are created, which are subject to two different threshold values, and reported in two panels of Fig~\ref{dirnetwork}. By comparing the two directed networks, we see overall associative relations among these 18 feature variables with depth. The feature variables of the Left are mutually associative, but this is not the case for the feature variables of the Right. There exist two separate cliques within the network in panel (A). Though these cliques become connected in panel (B), these two groups of feature variables are not mutually highly associated.

\begin{figure}
 \centering
   \includegraphics[width=6in]{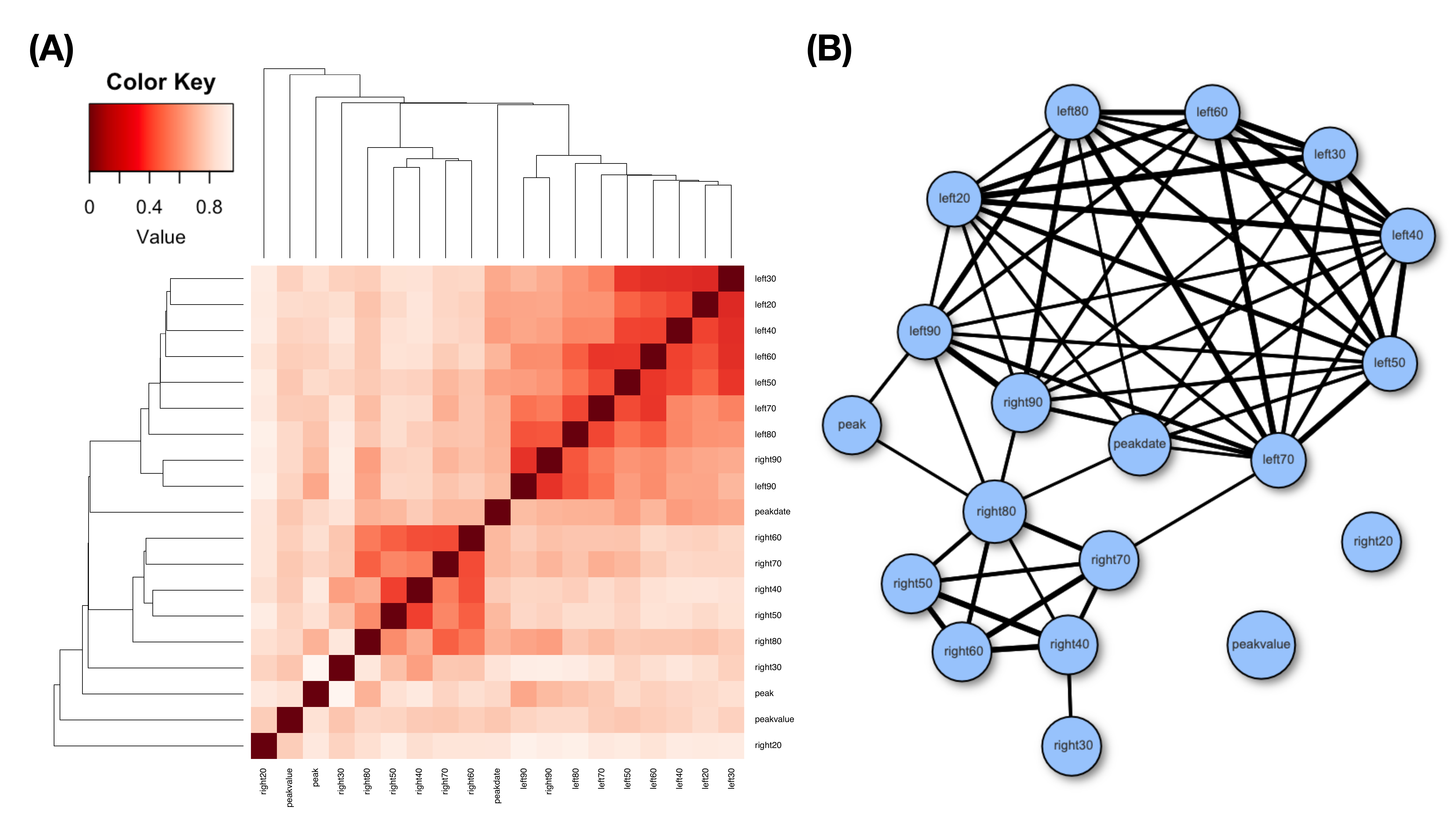}
 \caption{Associative heatmap (A) and network (B) of all features}
 \label{indirnetwork}
 \end{figure}

By making the simple average of the two directional associations, the conditional mutual information (MCE) is calculated as a nondirected association measurement of any pair of feature variables. Thus, another version of associative relations is represented through a heatmap of the MCE matrix and its corresponding network reported in Fig~\ref{indirnetwork}. Again, the feature variables on the Left form a complete network, while the feature variables on the Right don't.

Both Fig~\ref{dirnetwork} and Fig~\ref{indirnetwork} clearly show the asymmetric associations between and among the feature-variables of Left depicting the growth pattern and the feature-variables of Right depicting the decline. The implications of these pieces of pattern information should be understood as the constraints placed on the growth dynamics of this infectious disease within a closed system like Taiwan are more rigid than constraints placed on its counterpart of decline dynamics. This very interesting result seemingly tells us that there is not much room for growth patterns to vary when the infectious disease enters a pristine domain. That is, the shape of growth is more or less fixed except for slope and peak value. On the other hand, after reaching the peak value and beyond, the shape of decline can vary with large degrees of freedom. From the behavioral perspective, such a shape asymmetry between growth and decline seemingly indicates that behavioral impacts are expected to be also asymmetric as would be demonstrated in the next section.

\section{Major factors underlying dynamics of daily infection rate.}
In this section, we compute and discover the major factors underlying two curve characteristics of daily infection rate: peak value and curvature-at-peak. If the peak value is used to measure the strength of infection of Covid-19 within a district, then the curvature-at-peak measures the period of duration when this infectious disease can remain in full strength. Are features of the Left depicting growth or features of the Right depicting decline more associative to peak value or curvature-at-peak? We explore these two questions and report somehow surprising answers in this section.

\begin{figure}
 \centering
   \includegraphics[width=\textwidth]{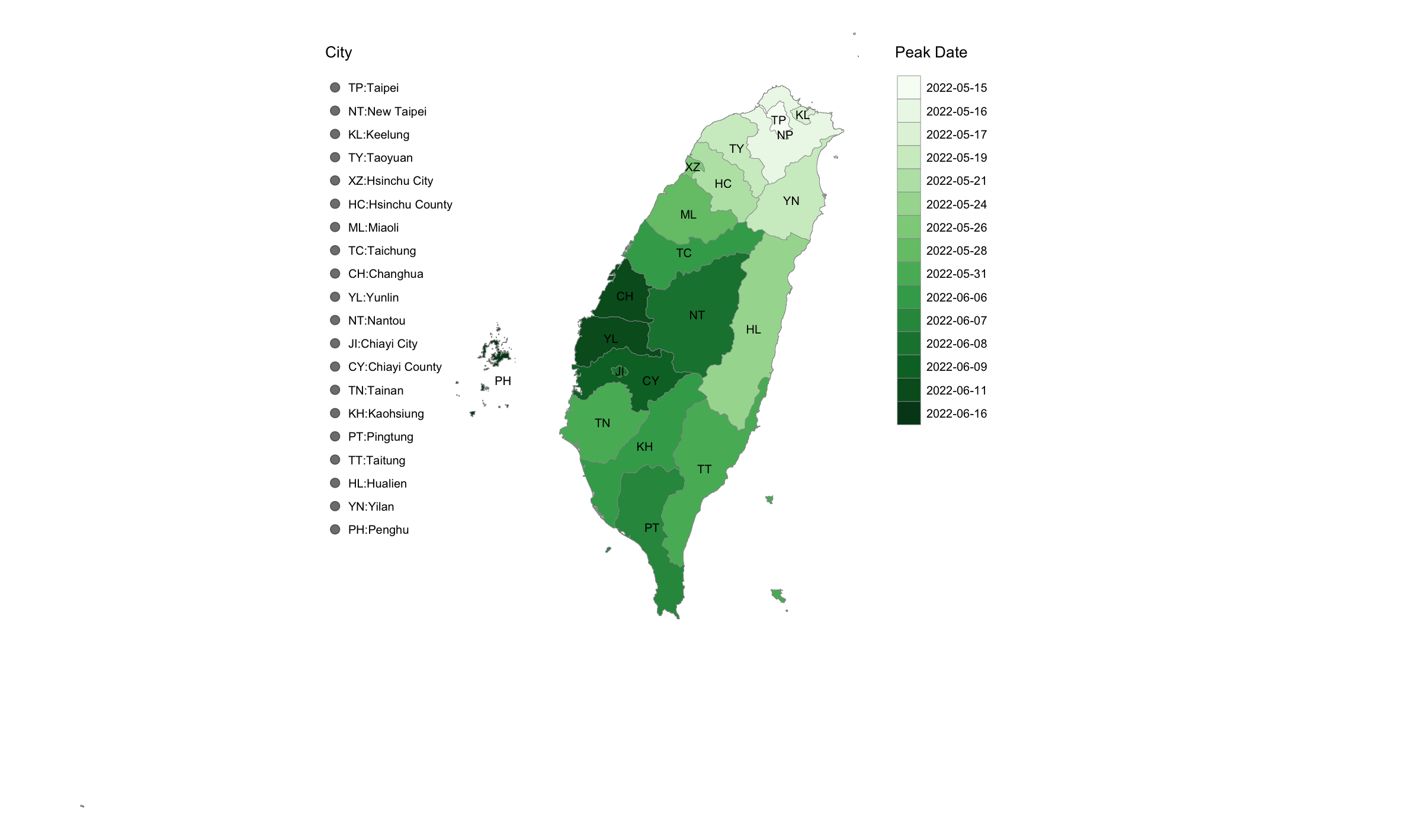}
 \caption{Visualization of the peak dates of all the cities in Taiwan}
 \label{TWdate}
 \end{figure}

But before exploring these two questions pertaining to the 84 districts, it might be helpful to visualize Covid-19 infection at the city scale in order to put all our discussions of spreading dynamics in a visible context. Fig~\ref{TWdate} shows the dates of peak across all cities in Taiwan. It is evident that this infectious disease started in Taipei (TP) and then spread to its suburban cities, such as New Taipei City (NT), Keelung (KL), and Taoyuan (KY). On the west side of Taiwan, which accommodates the island's majority of the population, the pathway of southward spreading indeed coincides with cities having major stations of the high-speed railway, such as Taichung (TC), Tainan (TN) and Kaohsiung (KS). Then, the spreading goes toward the suburbs of these three cities. In contrast, the east side of the island accommodates small cities of primary tourist attractions, such as Yilan (YN) and Hualien (HL). Their Covid-19 infections arrived rather early. Due to their populations being small in size, we do not consider these two cities in this study.

The Fig~\ref{TWdate} shows that Taipei has a smaller cumulative infection rate at its peak than its suburban cities of New Taipei, Keelung, and Taoyuan. Further, cities immediately south of these suburban cities are small in population size and less-densely populated. The varying cumulative rates in the three major cities: Taichung, Tainan, and Kaohsiung, are primarily due to averaging heterogeneous rates from their urban and suburban districts.

\begin{figure}
 \centering
   \includegraphics[width=\textwidth]{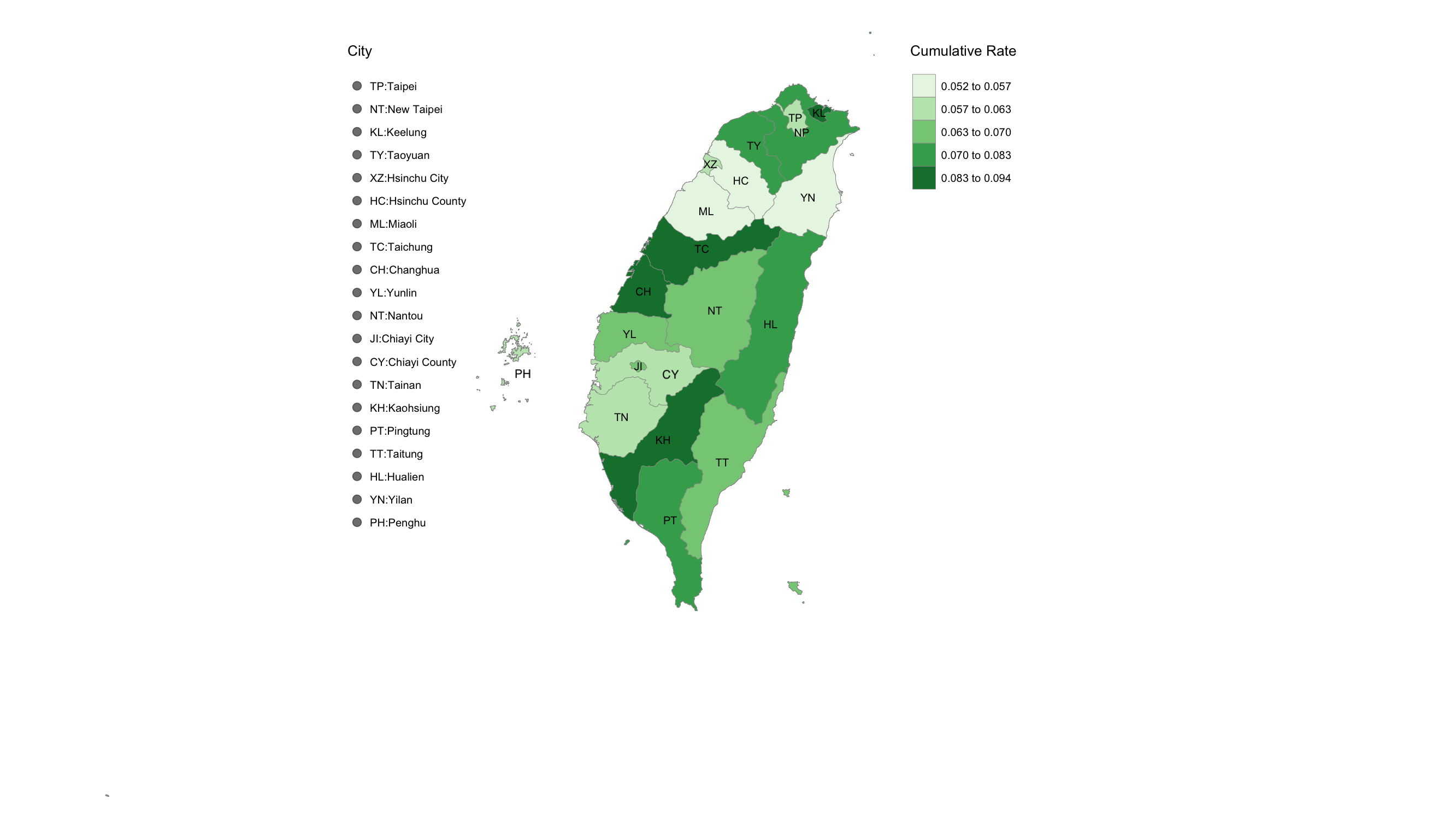}
 \caption{Visualization of the cumulative infection rates of all the cities at their peak dates in Taiwan}
 \label{TWrate}
 \end{figure}

\subsection{Exploring Peakvalue}
The height of the peak is the most eye-catching character pertaining to a district's curve of daily infection rate. It is natural to speculate that a district's growth pattern is likely to determine the height of its peak. Similar anticipation is that the height of its peak has the potential to greatly influence the decline pattern, on the other hand. To settle such natural speculation and anticipation, we conduct the CE-based evaluations to determine which covariate features or feature sets can influence the response variable: ``peakvalue''. The key part of CE calculations is reported in Table~\ref{peakvalue}.

Upon Table~\ref{peakvalue}, we see the top 5 ranked features and feature pairs. The most influential 1-feature turns out is ``right20'' at the very end of the daily infection rate curve. Their relationship can be reasoned as that the higher value of ``peakvalue'' will take a larger number of days to come down its $20\%$. The 2nd ranked 1-feature is ``left50'', which has a CE-drop significantly less than CE-drop of  ``right20''. Their relationship can be attributed to the highest growth slope being at its $50\%$ depicted by ``left50''. Due to being negative, the smaller ``left50'' is, the higher slope it will be, so is the ``peakvalue''. The 3rd ranked feature is ``right30''. The reasoning behind ``right30'' is about the declining speed at its $30\%$. The association between ``peakdate'' and ``peakvalue''' is reflecting one fact of spreading dynamics: the surge in Covid-19 infection rate started from districts in Northern Taiwan with a tendency depicted that the earlier surge is, the higher ``peakvalue'' is achieved.

In the 2-feature setting, we see the feature-pair (``peakdate'', ``right20'') achieves the lowest CE and at the same time reveals the ecological effect. That is,  ``right20'' and ``peakdate'' can concurrently be order-1 major factors underlying the dynamics of ``peakvalue''. In contrast, the 2nd ranked feature-pair  (``left70'', ``right20'') achieves a significantly larger CE than the CE of the feature-pair (``peakdate'', ``right20''), and their ecological effect is not revealed. Since their SCE-drop is smaller than the individual CE of ``left70'', which achieves less CE-drop than ``right20'' does. We conclude that the dynamics of ``peakvalue'' can be best described by a collection of two order-1 major factors:\{``peakdate'', ``right20''\}.

\begin{table}[h!]
\centering
\begin{tabular}{llllll}\hline
1-feature & CE & SCE-drop & 2-feature & CE & SCE-drop\\ \hline
right20&0.6507&0.5225&peakdate\_right20&0.3467&0.3039\\
left50&	0.8706&0.3025&left70\_right20&	0.4441&0.2065\\
right30&0.8739&0.2993&right90\_right20&0.4517&0.1989\\
peakdate&	0.8765&0.2966&left30\_right20&0.4611&0.1895\\
right70&0.8816&0.2916&left50\_right20&0.4675&0.1831\\
peak&1.0046&0.1685&left20\_left40&0.9042&0.0406\\\hline
\end{tabular}%
\caption{Top 5 and the bottom-ranked CEs of ``peakvalue'' as the response variable.}
\label{peakvalue}
\end{table}

We first take a look at the two order-1 major factors:\{``peakdate'', ``right20''\} separately. These two render two joint $4\times 4$ contingency tables reported in Table~\ref{Congpd20}, in which the ``R-0'' column denotes the missing values of ``right20''. It is striking to note from the first 0-column that there are 47 districts falling in the 0-category of ``right20'', which means these 47 districts have not yet declined and reached the $20\%$ of their ``peakvalue''. Beyond the first column, we see many zero cell counts of ``right20'' that make it have low CE. In contrast, the four rows of ``peakdate'' have even cell counts that make it have higher CE. The fact that this feature-pair \{``peakdate'', ``right20''\} achieves the ecological effect renders their  $4\times 16$ contingency table will contain many cells with zero-counts in order to be concurrently present as two order-1 major factors.

\begin{table}[h!]
\centering
\begin{tabular}{l|lllll|lllllllllll}\hline
pv/R-pd&R-0&R-1&R-2&R-3&R-4&pd-1&pd-2&pd-3&pd-4\\ \hline
1&13& 6 & 0 &0& 5 & 3 &6&6&9\\
2&29&3& 0 &  0& 0 &1 &3&3&25\\
3&2&  8& 4 &6 & 3 &8&11&3&3\\
4&3&  0& 0 &0 & 0 &0&2&0&1\\
\hline
\end{tabular}%
\caption{Joint two $4\times 4$ contingency tables of  ``right20'' and ``peakdate''  vs ``peakvalue''. The ``0'' means ``NA'' for ``right20''.}
\label{Congpd20}
\end{table}

\subsection{Exploring curvature-at-peak}
The curvature at the robust-peak $t_{0}$ is a key characteristic of any district-specific smoothed curve of daily infection rate. A large curvature corresponds to a small ``right90'' value and indicates a sharp growth coupled with a steep decline. In contrast, a small value of curvature-at-peak strongly indicates that the district's infection rate is somehow maintained by the ``full infection force'' for a lengthened period centered at robust-peak $t_{0}$. Such a somehow leveling-off pattern is not ideal for the community within the district.

We explore the dynamics of curvature-at-peak $t_{0}$ by using ``right90''  as a response variable. Its close counterpart ``left90''  will be excluded as a covariate feature. We report the top 5 and bottom 2 ranked CEs in 1-feature and 2-feature settings. In the 1-feature setting, it is somehow surprising to see that the feature``left80'' ranked at the top, while feature ``right80'' ranked 5th. There is a rather significant difference in their CE-drops. It is not surprising that ``left80'' is ranked at the top because of its high association with ``left90'', which is nearly equal to ``right90''. However,  what is surprising is that feature ``right80'' is ranked 5th after ``left50''. This asymmetric result further indicates another aspect of the difference between growth and decline patterns in curves of daily infection rate.

In the 2-feature setting, we see that the feature-pair (``left80'', ``right80'') achieves the lowest CE, but fails to reveal the ecological effect. We joint two $4\times 4$ contingency tables: $C[``left80''-vs-``right90'']$ and $C[``right80''-vs-``right90'']$, and report them in Table~\ref{Cong8080} to demonstrate the dominant effects of ``left80'' in reducing uncertainty of ``right90'', while ``right80'' doesn't do much in extra.

\begin{table}[h!]
\centering
\begin{tabular}{llllll}\hline
1-feature & CE & SCE-drop & 2-feature & CE & SCE-drop\\ \hline
left80&	0.5046&0.7138&left80\_right80&0.2440&0.2606\\
left70&	0.6386&0.5798&left80\_right20&	0.3071&0.1975\\
left60&	0.6957&0.5228&left80\_right70&	0.3307&0.1739\\
left50&	0.7128&0.5056&left80\_right50&	0.3480&0.1566\\
right80&0.7644&0.4540&left80\_peakdate&	0.3651&0.1395\\
peakvalue&0.9991&0.2193&peakvalue\_right40&0.8287&0.1474\\
right30&1.1000&0.1014&right40\_right30	&0.9016&0.0746\\\hline
\end{tabular}%
\caption{Top 5 and bottom 2 ranked CEs of ``right90'' as the response variable for curvature.}
\label{right90}
\end{table}

\begin{table}[h!]
\centering
\begin{tabular}{l|llll|llllllllllll}\hline
R90/L-R&L-1&L-2&L-3&L-4&R-1&R-2&R-3&R-4\\ \hline
1&4& 2& 1 & 34 &26&11&2&2\\
2&0& 3& 10 & 0 &0&0&10&3\\
3&0& 10& 1&  0 &0&2&2&7\\
4&11& 4 & 0 & 2&11&1&5&0\\
\hline
\end{tabular}%
\caption{Joint two $4\times 4$ contingency tables of ( ``left80'', ``right80'') vs ``right90'' table.}
\label{Cong8080}
\end{table}

\section{Geographic and social-economic effects.}
In this section, we look into whether the growth and decline patterns of daily infection rate reveal signs of geographic and social-economic differences with respect to 84 districts in the largest seven cities. To these two effects, we introduce two new binary features: North-vs-South for geographic effect and urban-vs-suburban for social-economic effect. All involving districts are encoded with North or South categories according to the cities to which they belong: Taipei (TP), New-Taipei city (NT), Keelung (KL), and Taoyuan (TY) for North and Taichung (TC), Tainan (TN) and Kaohsiung (KH) for the South. Likewise, each district is assigned an urban or suburban category. These two binary features would be used as response variables in our CE computations, respectively.

\subsection{Exploring geographic effects}
As for the growth pattern, we apply K-means (K=4) clustering algorithm to fuse a segment of serial-dependent features into one single fused feature. This is a way of avoiding the effect of the curse of dimensionality. For instance, a district's 5dim vector (left30, left40, left50, left60, left70) will be assigned to a category according to which one of the four clusters it falls into. We denote this clustering-based feature of growth as ``left30to70''. Likewise, we have two more features of growth named: ``left30to50'' and ``left30to60''. As for the decline patterns, we also have three fused features of decline named as:``right30to50'', ``right30to60'', and ``right30to70''. These 6 fused features and their pairs are used to reduce the uncertainty of the two response variables. Their CEs and SCE-drops are reported in Table~\ref{NSeffect} and Table~\ref{Urbaneffect}, respectively.

Upon Table~\ref{NSeffect}, the fused feature of growth ``left30to70'' achieves the lowest CE followed by ``left30to60'' and ``left30to50''. Their CE-drops are five times or more of the CE-drops of the three fused features of decline: ``right30to70'', ``right30to60'' and ``right30to50''. By interpreting such CE-drops as the mutual information between each individual fused feature and the response geographic feature: North-vs-South, we conclude that geographic effects are evidently manifested through growth patterns rather than through decline patterns.

As for the 2-feature setting, we see that the pair of fused features ( left30to70, right30to70) achieved the lowest CE while the ecological effect is observed. Since the SCE-drop: 0.1017, is larger than the CE-drop of ``right30to70''. This result assures that the combined growth and decline pattern indeed can better reveal the geographic effect on dynamics underlying the Covid-19 daily infection rate. See the strong evidence contained in the contingency table of ( left30to70, right30to70) vs North-vs-South in Table~\ref{Conggeo} by having the majority of columns with very low or even zero CEs. We also see similar, but fewer effects through other combinations of growth and decline patterns. In contrast, growth-and-growth and decline-and-decline combinations have no significant effects beyond their individual effects.

\begin{table}[h!]
\centering
\begin{tabular}{llllll}\hline
1-feature & CE & SCE-drop & 2-feature & CE & SCE-drop\\ \hline
left30to70 &	0.27619&0.4130& left30to70\_right30to70&0.1744& 0.1017\\
left30to60	&0.2955& 0.3936& left30to50\_right30to70 &0.1819&0.1167\\
left30to50 &	0.2986&0.3905&left30to70\_right30to60&0.2100&0.0661\\
right30to70 &	0.6059&0.0832&left30to60\_left30to70	&0.2699&0.0062\\
right30to60 &	0.6144&0.0747&right30to50\_right30to70	&0.5269&0.0789\\
right30to50 &	0.6300&0.0591&right30to50\_right30to60 &	0.6104&0.0039\\\hline
\end{tabular}%
\caption{Testing North-vs-South for geographic effect.}
\label{NSeffect}
\end{table}

\begin{table}[h!]
\centering
\begin{tabular}{lllllllllllllllll}\hline
N-S/L\_R &1-1&1-2&1-3&1-4&2-1&2-2&2-3&2-4&3-1&3-2&3-3&3-4&4-1&4-2&4-3&4-4\\ \hline
North&4&0&5&11&0&0&0&0&2&3&12&5&0&0&1&0\\
South&1&2&0&0&3&5&6&1&1&0&1&1&4&2&8&1\\\hline
\end{tabular}%
\caption{Contingency of ( left30to70, right30to70) vs North-vs-South table for North-South geographic effect.}
\label{Conggeo}
\end{table}

\subsection{Exploring social-economic effect}
Taipei is the capital city of Taiwan. From many aforementioned aspects, such as housing prices and availability of open public spaces among many others, its 12 districts enjoy higher social-economic status than districts in New Taipei city, Keelung and Taoyuan. Therefore, we encode these 12 districts as urban, while the 36 districts in the three cities surrounding Taipei as suburban. In contrast, Taichung, Tainan, and Kaohsiung consist of urban and suburban districts, since their current city administration is recently expanded by including districts from counties under the same name.

Upon Table~\ref{Urbaneffect}, we discover several highly striking findings. In the 1-feature setting, all individual fused features show a rather limited urban-vs-suburban effect. However, we see outstanding interacting phenomena in the 2-feature setting. The pair (left30to60, right30to70) achieves an SCE-drop:0.1504, which is more than seven times of CE-drop of ``right30to70''. This effect is also visible in Table~\ref{Urbantable} by having the majority of columns with very low or even zero CEs. This interacting effect strongly indicates that the urban-vs-suburban-based social-economic effect can only be better revealed through the combined growth-and-decline patterns. The other two pairs (left30to70, right30to50) and (left30to50, right30to70) show similar strong interacting effects. In comparison, pairs of two growth features or decline features clearly do not have such effects at all.

These outstanding interacting effects of growth and decline patterns strongly imply that curves of daily infection rates derived from urban districts are fundamentally distinct from curves derived from suburban districts. Such results are coherent with our observations that urban curves are essentially lower in peak value and smaller in curvature-at-peak.

In summary, we successfully evaluate geographic and social-economic effects underlying the spreading dynamics of Covid-19 daily infection rates. In this section, we also demonstrate a way of avoiding the curse of dimensionality via clustering algorithms.

\begin{table}[h!]
\centering
\begin{tabular}{llllll}\hline
1-feature & CE & SCE-drop & 2-feature & CE & SCE-drop\\ \hline
left30to60&0.6338&0.0359&left30to60\_right30to70	&0.4833&0.1504\\
left30to50&0.6387&0.0310&left30to70\_right30to50&0.4864&0.1590\\
left30to70&0.6454&0.0243&left30to50\_right30to70 &	0.4943&0.1444\\
right30to70&0.6507&0.0190&left30to60\_right30to60&0.5731&0.0607\\
right30to50&0.6510&0.0187&left30to50\_left30to70&0.6326&0.0061\\
right30to60&0.6616&0.0081&right30to50\_right30to60&0.6503&0.0006\\\hline
\end{tabular}%
\caption{Testing urban-vs-suburban for social-economic effect.}
\label{Urbaneffect}
\end{table}

\begin{table}[h!]
\centering
\begin{tabular}{llllllllllllllll}\hline
 U-S/L\_R &1-1 &1-2 &1-3 &1-4& 2-1& 2-2& 2-3 &2-4 &3-1 &3-2 &3-3 &3-4 &4-1 &4-2 &4-3\\
urban      &  1  & 1 &  2 &  1  & 0  & 1  & 7  & 7  & 2 &  4 & 1  & 0 &  1 &  1 &  2\\
suburban & 0 &  0  & 2  & 0  & 9  & 4&  11  &10   &0  & 0 &  5 &  1 &  2  & 1 &  3\\
\hline
\end{tabular}%
\caption{Contingency table of ( left30to60, right30to70) vs urban-vs-suburban for testing urban-vs-suburban for social-economic effect.}
\label{Urbantable}
\end{table}

\section{Growth and decline similarity detailed.}
Based on the non-directed and directed networks among all feature-variables presented in Fig~\ref{dirnetwork} and Fig~\ref{indirnetwork}, we see the feature-variables of Left are highly associative with each other, while the feature-variables of Right are clearly less associative. These two groups of feature variables are neither highly associative, except the pair ``left90'' and ``right90''. Further, the CEs computed for testing the geographic effect of North-vs-South also strongly indicate that those feature variables on the Left are more critical than the feature variables on the Right. Based on this knowledge, we evaluate the degrees of similarity among all districts and age groups based on feature variables of Left and Right, respectively. Euclidean distance would be used as a similarity measure, since this similarity measure would mix all feature variables together. It is not surprising to expect neither entirely blurred nor clear-cut resultant patterns regarding similarity among district- or age-group-specific curves of daily infection rates. These are reasons why we compute the degree of similarity based on the feature variables of the Left and feature variables of the Right, respectively.

\subsection{Geographic effect via similarity among districts.}
In this subsection, we consider similarity on the district scale of the seven cities. Each district is given a three-letter code name. A two-capital-letter code is assigned to each city as shown in Fig~\ref{TWdate} and Fig~\ref{TWrate}, for instance ``TP'' for Taipei and ``NP'' for New Taipei. This two-capital-letter code is followed by a small letter ranging from ``a to l'' indicating one of the city's 12 districts.

Among these 84 code-named districts, their degrees of similarity based on feature variables of Left are evaluated via Euclidean distance. The Hierarchical clustering (HC) algorithm with the Ward-d2 module is employed to build a clustering tree. This HC-tree has 84 tree leaves located and grouped with respect to branches. A set of cod-names sharing one tree branch are intuitively taken as being similar to a degree specified by the tree level where this branch is located. That is, if two code-named leaves are found together within a branch at the bottom level, then this pair of code-named districts are very similar. The degree of this visible similarity can be further numerically evaluated upon this HC-tree's tree geometry. This evaluation is carried out through a binary coding scheme proposed and developed in \cite{chou}.

\begin{figure}
 \centering
   \includegraphics[width=6in]{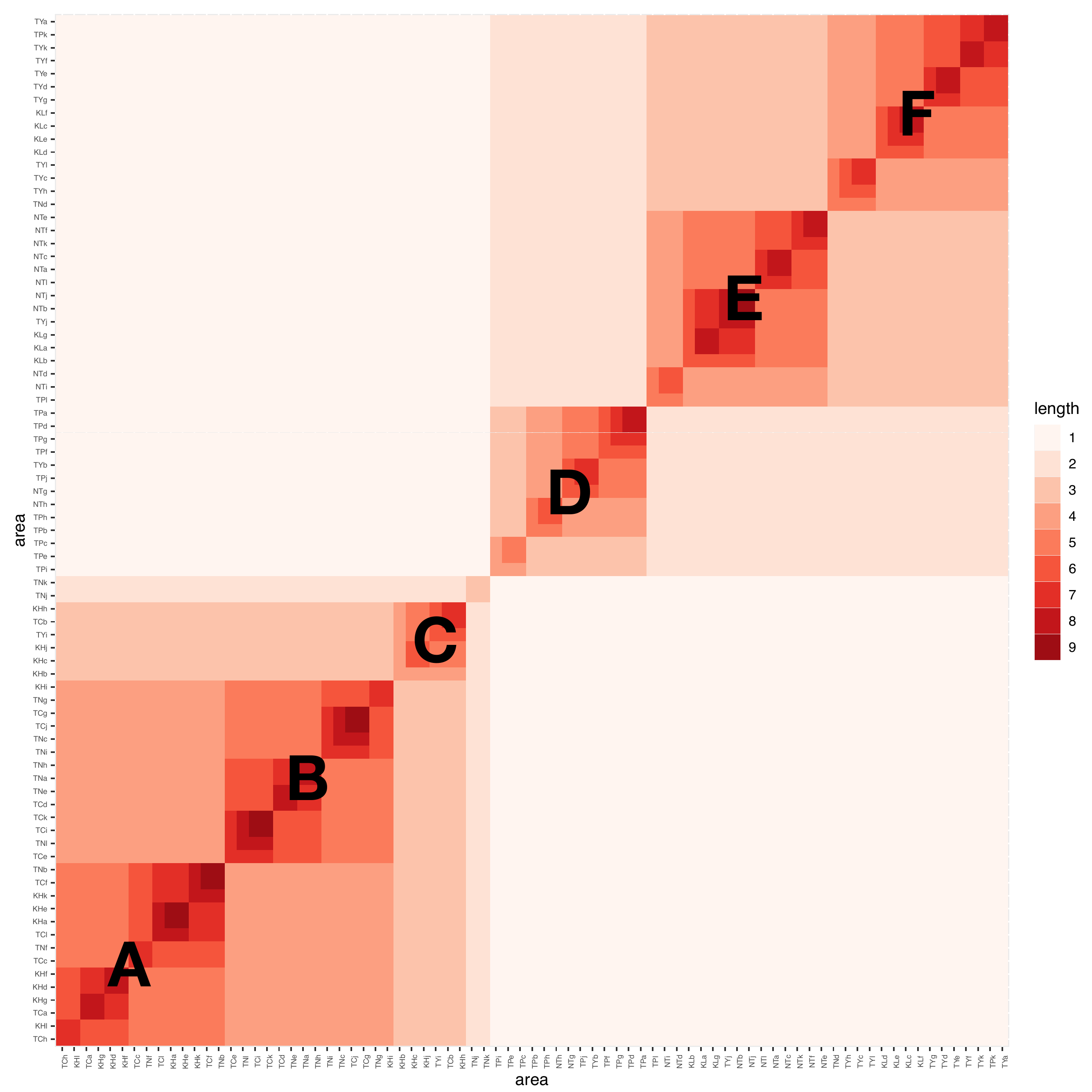}
 \caption{Growth similarity among all districts.}
 \label{leftallcity}
 \end{figure}

The HC-tree is binary in the sense that one branch is spilt into two subbranches at its internal node. Therefore, a simple binary coding scheme: left for 0 and right for 1, can be used to encode each tree leaf (or code name) by going from the top level to the bottom level of this tree. This binary coding scheme is simply a way of locating each tree leaf by using a segment of binary codes. Two tree leaves sharing a lengthier binary coding segment are more similar to each other than two tree leaves sharing a shorter binary coding segment. That is, this length of common coding segment is a precise measure of the degree of similarity between any two tree leaves. We build a heatmap based on such a tree-distance matrix as shown in Fig.~\ref{leftallcity}. Code-named districts along the matrix's row and column axes are arranged with respect to the HC-tree.

This heatmap reveals evident block patterns of two scales: big and median, in Fig.~\ref{leftallcity}. There are 8 median blocks marked with capital letters from ``A'' to ``H''. There are two big blocks. The upper-right big block consisting of \{D, E, F,G, H\} median blocks precisely contains all districts belonging to the four cities in the North: Taipei (TP),  New Taipei (NP), Taoyuan (TY) and Keelung (KL). Among these 5 median blocks, except the median-block-D exclusively consisting of 8 out of 12 districts of Taipei, district memberships in the rest of 4 median blocks are rather mixed among districts primarily from New Taipei (NP), Taoyuan (TY) and Keelung (KL).

In contrast, the lower-left big block consists of 3 median blocks:\{A, B, C\}. The smallest median block C has 4 member districts from Kaohsiung exclusively. Nonetheless, it is interesting to note that the median-block-A and -B are neither well mixed with district memberships belonging to Taichung (TC), Tainan (TN), and Kaohsiung (KH). The median-block-A contains nearly all districts from Tainan (TN) and some districts from Taichung (TN). In contrast, the median-block-B is consisting of districts mainly from Taichung (TC) and Kaohsiung (KH). However, Tainan (TN) is geographically located between Taichung (TC) and Kaohsiung (KH).

This significant and evident separation between the North and South is primarily based on feature variables of the Left for the growth pattern of the curve of daily infection rate. Such results echo the testing results of geographic effects reported in the previous section.

\begin{figure}
 \centering
   \includegraphics[width=6in]{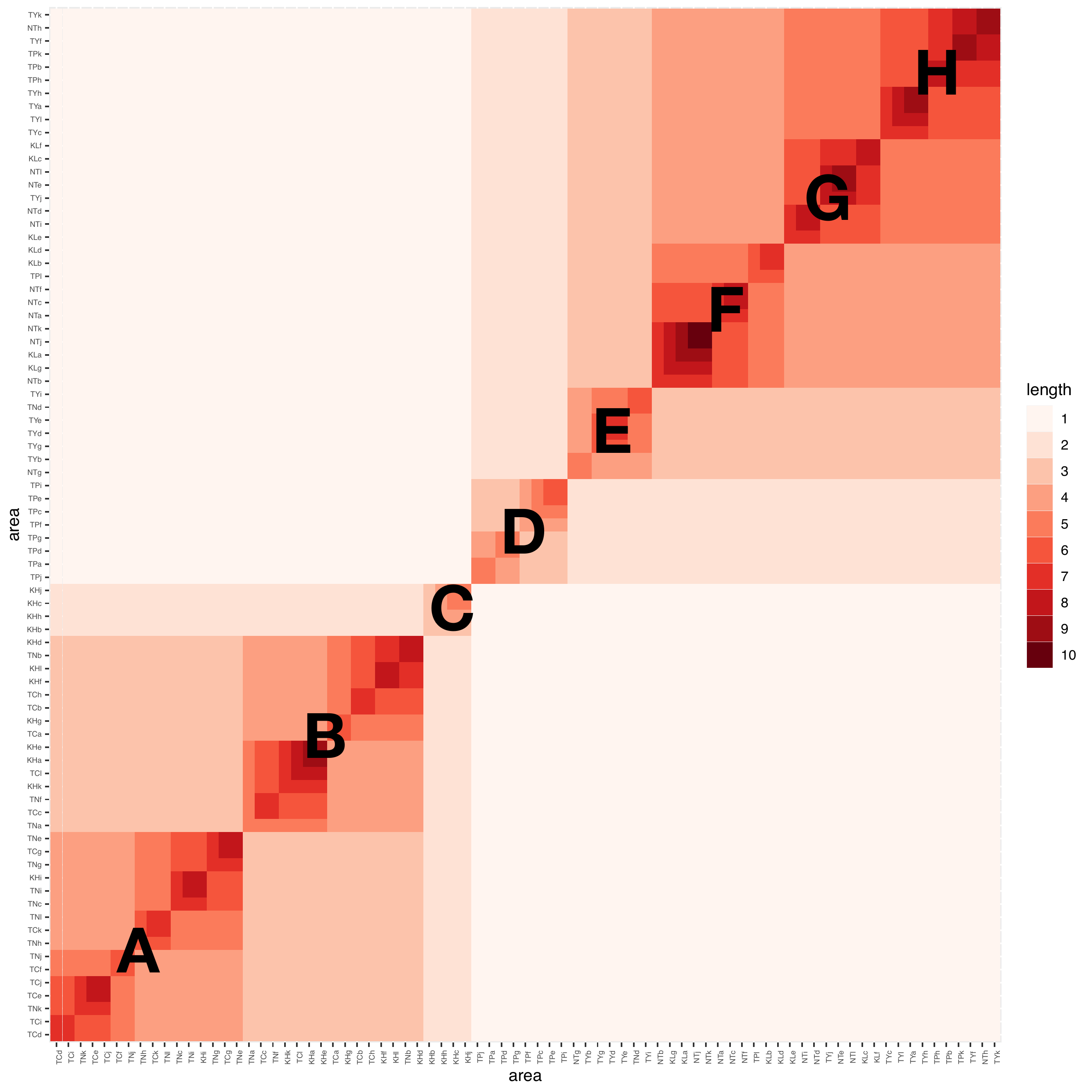}\\
 \caption{Decline similarity among all districts.}
 \label{rightallcity}
 \end{figure}

Likewise, we explore the similarity from the decline perspective by using these 84 districts' measurements of feature variables of the Right. We also build an HC-tree with 84 tree leaves at the district level. Upon this HC-tree, we again apply the same binary coding scheme to construct an $84\times 84$ tree-distance matrix.  In Fig.~\ref{rightallcity}, we report the tree-distance-based heatmap with row- and column-axes being arranged according to the HC-tree. Again we see the clear separation between the North and South. The big block of the North consists of median blocks:{D, E, F}, while the big block of the South consists of median blocks:{A, B, C}. It is interesting to see that median block D consists of 13 districts: 11 districts of Taipei (TP) and 2 districts of New Taipei city (NT).

Again an interesting pattern is seen by comparing district memberships between median-block-A and median-block-B. The majority of districts of Tainan (TN) are in median-block-B, which also contains some districts of Taichung (TC). In comparison, district memberships of median-block-A are nearly exclusively from Taichung (TC) and Kaohsiung (KH).  It seems that the districts of Taichung (TC) are partly similar to Tainan (TN) and partly similar to Kaohsiung (KH) from the decline perspective. However, geographically, Tainan (TN) is in between Taichung (TC) and Kaohsiung (KH).

In summary, from both growth and decline perspectives, we successfully present multiscale patterns of similarity among 84 city-districts through multiscale block patterns found on two heatmaps of tree distances or lengths of shared coding segments. Such results are coherent with results regarding geographic effort. Further, our representation of degrees of similarity derived from HC-trees indeed provides a new way for pattern recognition. It is essential to note that such pattern information is visible and interpretable.

\subsection{Social-economic effect via similarity among age groups}
From a behavioral perspective, similarity among curves of daily infection rate pertaining to age groups intuitively should reflect social-economic effect, while the similarity among districts should reflect geographic effect. This intuition is based on the fact that different age groups are linked with distinct aspects of social-economic aspects and reveal their heterogeneous characteristics. For instance, different school age groups link to different school districts that implement distinct Covid-19 regulations. Young adult and middle age groups belonging to distinct districts need to travel through different means to arrive at the same workplaces and locations of companies, while different senior age groups link to different available public open spaces and facilities with heterogeneous conditions. Hence, the social-economic effect via urban-vs-suburban statuses is expected to be manifested through similarity pertaining to age groups.

Four age groups are considered and identified via the 4th digital code: 1 for 0-19; 2 for 20-34; 3 for 35-54 and 4 for 55+, attaching to the 3 letter-code of the district. In this subsection, we consider these 4 age groups belonging to the 12 districts of Taipei (TP) and 12 districts of New Taipei City (NT). Like in the previous subsection, we evaluate the similarity among these 96 curves of daily infection rate.

\begin{figure}
 \centering
   \includegraphics[width=6in]{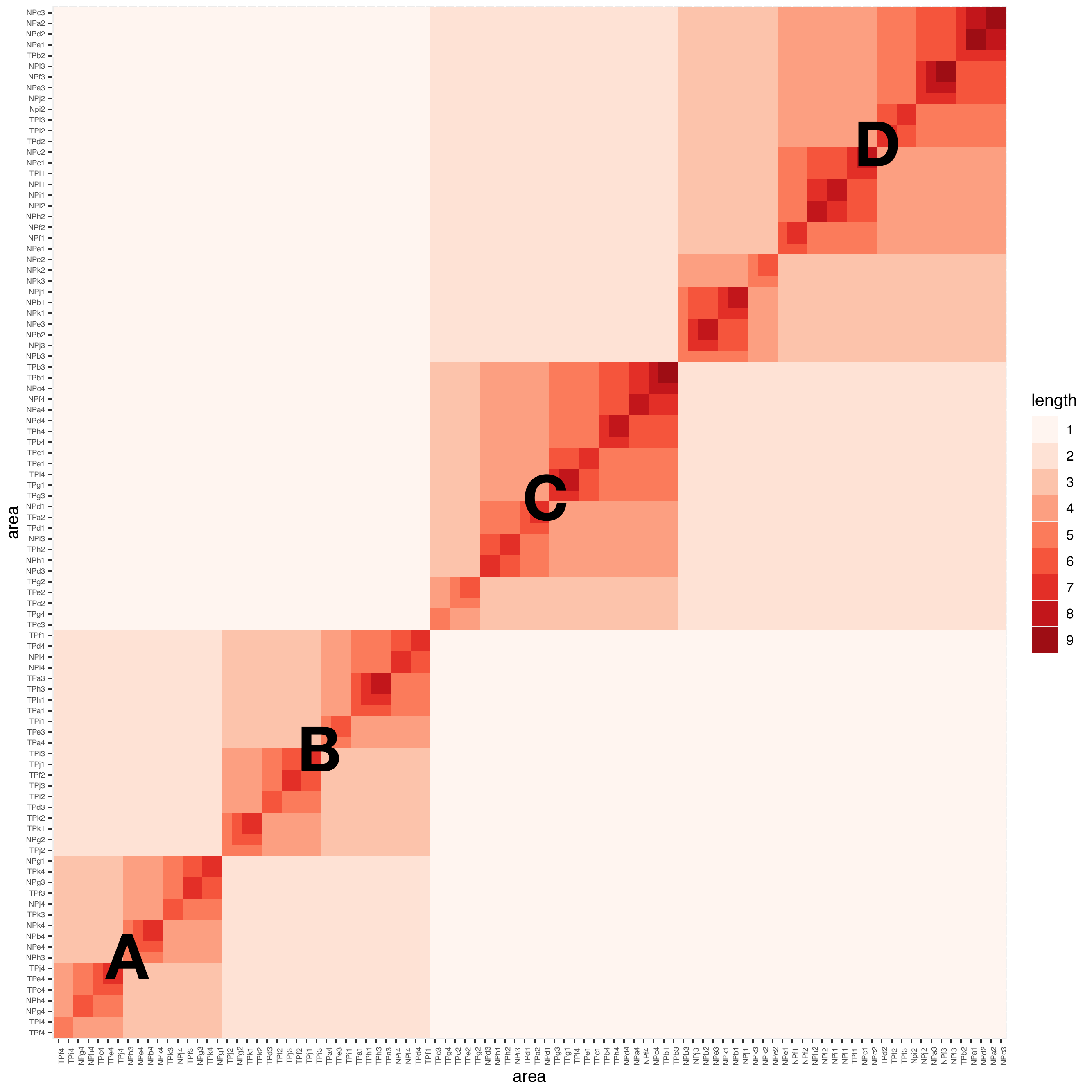}\\
 \caption{Growth similarity among age groups in two Taipei and New-Taipei cities.}
 \label{leftage}
 \end{figure}

From the growth perspective pertaining to features of Left, the heatmap of lengths of the coding segment, which is constructed by implicitly superimposing the HC-tree with 96 leaves onto its row- and column-axes, shown in Fig~\ref{leftage} reveals visible multiscale block patterns. This heatmap reveals visible multiscale block patterns. For expositional simplicity, we mark two median blocks in each of the two big blocks with capital letters: A to D, respectively. The median-block-A and median-block-B within the lower-left big block are evidently divided. The code-IDs of members in median-block-B are primary age groups 1, 2, and 3 belonging to TP, contrastingly code-IDs of members in median-block-A are primary age groups 4 belonging to NT and TP. Further, though, the code-IDs of members in median-block-C are mixed in terms of age groups and TP-vs-NT, contrastingly code-IDs of members in median-block-D are primary age groups 1, 2, and 3 belonging to NT.

\begin{figure}
 \centering
   \includegraphics[width=6in]{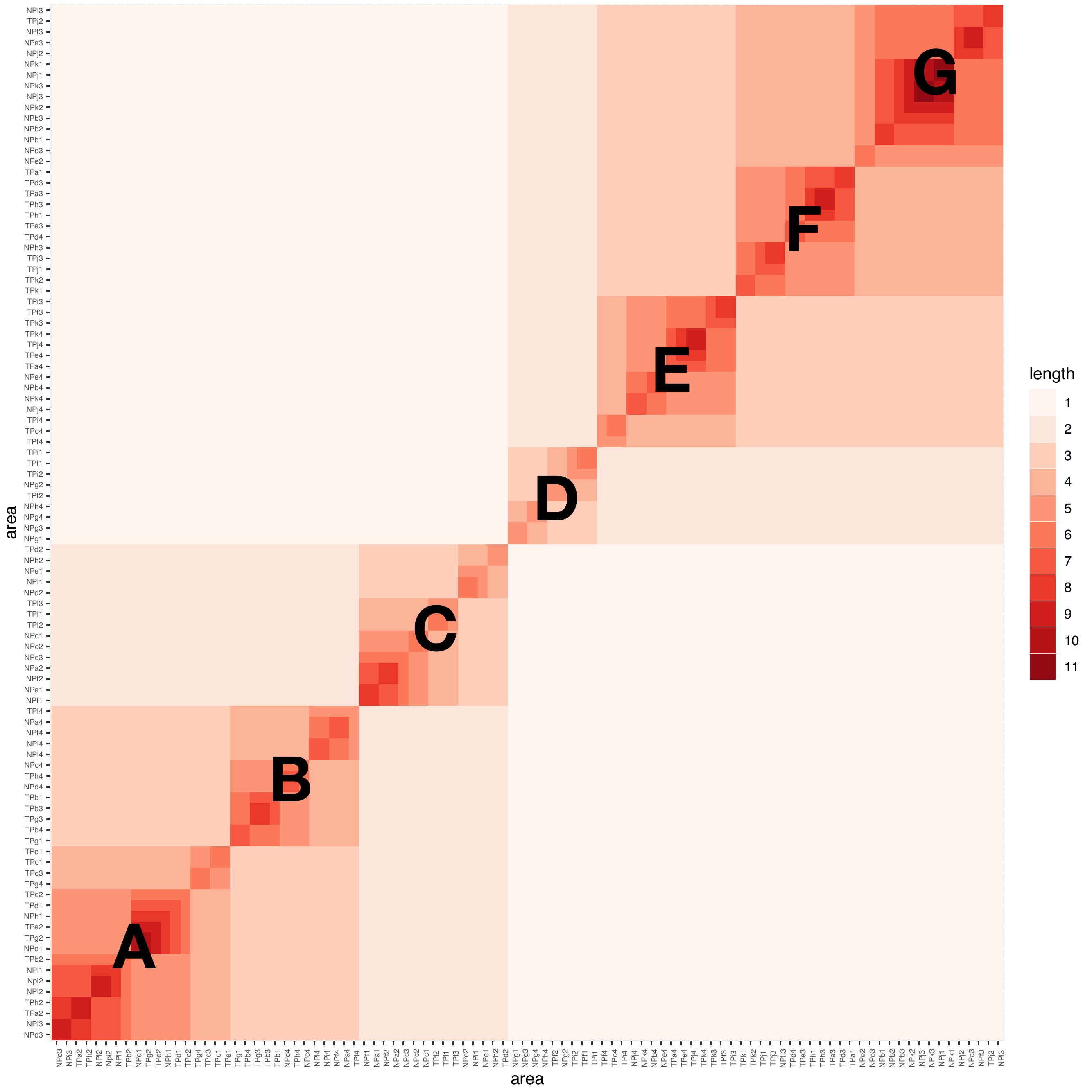}\\
 \caption{Decline similarity among age groups in two Taipei and New-Taipei cities.}
 \label{rightage}
 \end{figure}

From the decline perspective pertaining to features of Right, the heatmap of lengths of the coding segment is likewise constructed by implicitly superimposing the HC-tree with 96 leaves onto its row- and column-axes, shown in Fig~\ref{rightage}. This heatmap reveals evident block structures with two scales. The big-block scale: lower-left and upper-right, shows the most evident separation by design. Also, we mark 3 median blocks in the lower-left big block, and 4 median blocks in the upper-right big block by capital letters: A to G, respectively. The median-block-A consists of 15 members. These 15 code IDs are primarily of age groups 1 and 2 coming from districts in both TP and NT,  so are 17 members of the median-block-C. In contrast, the 13 members of median-block-B are primarily of age-group 4, and so are the 14 members of median-block-E. The members of median-block-F and -G are mixing with age groups 1, 2, and 3 from both cities.

In summary, as intuitively expected from the social-economic effect perspective, the separation of suburban NT against urban TP is seen with respect to growth patterns among age groups, but not evident in terms of decline patterns.

\section{Conclusions}
This paper studies spreading characteristics and human behavioral effects embedded within the dynamics of Covid-19 infection within Taiwan. The study period is designed and chosen when this island nation was as pristine as well as a nearly close domain to this infectious disease. Structured data derived from 84 curves of daily infection rates pertaining to 84 districts of 7 cities is analyzed via Theoretical Information measurements to discover major factors determining curves' two key characteristics: peak values and curvature-at-peak. Further, human behavior oriented geographic and social-economic effects are linked to growth and decline patterns of curves' shape-formations. Furthermore, we apply the hierarchical clustering algorithm on growth and decline data of 84 curves and 96 age-group-specific curves respectively to construct tree-distance based heatmaps as an alternative way of confirming both North-vs-South and Urban-vs-Suburban effects.

Here we briefly summarize our contributions of this paper. The first contribution is the successful extraction of 18 features from unstructured curves of daily infection rates pertaining to two scales: 1) the city's district; 2) the district's age group. Such structured data format enables us to collectively characterize the growth and decline patterns of all curves pertaining to the two scales. This extraction becomes a natural way of analyzing functional data.

The second contribution is the analytic presentations of asymmetry of between the curve's growth and decline patterns by employing conditional entropy based major factor selection protocol. Through the ``peakvalue'' and ``curvature-at-peak'' as two separate response variables, we discover different collections of major factors underlying both dynamics, respectively. The ``peakvalue'' is highly associated with features of the Left that depict the growth patterns, while the ``curvature-at-peak'' is highly associated with features of the Right that characterize the decline patterns. From these two computational results, we can go beyond the verbal descriptions of asymmetry regarding visible growth and decline patterns. From a computing and science perspective, this major factor selection protocol demonstrates how to peek into complex dynamics without assuming man-made structures.

The third contribution is the established geographic effect via a binary encoding on all districts: North-vs-South, and the social-economic effect via another binary encoding on all districts: urban-vs-suburban. These results are significant because neither individuals' geographic nor social-economic data is being used. Both behavioral oriented effects are further confirmed by making use of district-specific and age-group-specific curve-similarity with focal regions of growth and decline, respectively. It is also worth noting that our coding-length-based tree-distance heatmap indeed provides a new way of representing a resultant hierarchical clustering tree for better visualization.

Finally, we synthesize these three contributions into one take-home message: human behaviors do have computable profound impacts on the spreading dynamics of Covid-19 infection. We hope this message from Taiwan could add a brand-new perspective to studying Covid-19 and many other infectious diseases' spreading dynamics.

\bibliographystyle{unsrt}

\end{document}